\documentclass[journal]{IEEEtran}
\usepackage{amsmath,amsfonts,amssymb}
\usepackage{algorithmic}
\usepackage{algorithm}
\usepackage{array}
\usepackage[caption=false,font=normalsize,labelfont=sf,textfont=sf]{subfig}
\usepackage{textcomp}
\usepackage{stfloats}
\usepackage{url}
\usepackage{verbatim}
\usepackage{graphicx}
\usepackage{cite}
\usepackage{booktabs}
\usepackage{multirow}
\usepackage{wasysym}

\newcommand{\bpp}{\text{bpp}}

\newcommand{\EMPTYcircle}{\Circle}
\newcommand{\etal}{\textit{et al.}}

\begin{document}

\title{
Adapting Diffusion Language Models for Lossless Pixel-Level Image Transmission
}

\author{Tianqi Ren, Rongpeng Li, Xianfu Chen, Yingyu Li, and Zhifeng Zhao\\
\thanks{T. Ren and R. Li are with College of Information Science and Electronic Engineering, Zhejiang University, Hangzhou 310027, China (email: \{rentianqi, lirongpeng\}@zju.edu.cn). X. Chen is with Shenzhen CyberAray Network Technology Co., Ltd, Shenzhen 518000, China (email: xianfu.chen@ieee.org). Y. Li is with School of Mechanical Engineering and Electronic Information, China University of Geosciences, Wuhan 430074, China (email: liyingyu29@cug.edu.cn). Z. Zhao is with Zhejiang Lab, Hangzhou 311121, China (email: zhaozf@zhejianglab.org).

}
}
\maketitle

\begin{abstract}
Lossless pixel-level image transmission is a fundamental regime beyond semantic communications, because exact recovery requires both accurate symbol probability modeling and reliable delivery over noisy channels.
This paper proposes DDM-SSCC, a discrete-diffusion-model-based separate source-channel coding framework for lossless image transmission.
Different from raster-order autoregressive coding, the proposed source codec adapts a diffusion language model to pixel-token restoration and performs synchronized reverse arithmetic coding under bidirectional attention, allowing multiple masked tokens to be coded within one reverse denoising step.
This progressive restoration process also yields a more favorable source representation for noisy transmission, since newly restored tokens can serve as bidirectional context in subsequent denoising steps.
To bridge the gap between generation-oriented masked denoising and lossless arithmetic coding, we further introduce a Halton-guided denoising order, a mask-ratio-aware cosine schedule, and a lightweight temperature calibration module.
These designs respectively improve spatial coverage, adapt the denoising pace to context reliability, and calibrate the probability tables used by arithmetic coding.
Experiments on CIFAR10, DIV2K-LR-X4, and Kodak over additive white Gaussian noise and Rayleigh fading channels show that DDM-SSCC achieves better exact-recovery performance than representative lossless and semantic communication baselines, while ablation studies verify the effectiveness of the proposed denoising order, schedule, and calibration modules.
\end{abstract}


\section{Introduction}

\IEEEPARstart{R}{}ecently, semantic communications have manifested strong effectiveness for distortion-oriented delivery.
However, the underlying deep joint source-channel coding (JSCC) inevitably incurs pixel-wise reconstruction errors \cite{bourtsoulatze2019deep,tong2025sparsesbc,zhang2023deepma,jia2023lightweight}.
This limitation becomes critical in fidelity-sensitive scenarios such as remote medical imaging, industrial inspection, and scientific image delivery, where the receiver must recover the original image exactly.
By contrast, lossless pixel-level transmission provides a necessary probabilistic interface to the raw image signal, and serves as a foundation for future patch-level or higher-level semantic representations.
Recent progress in large-model-based compression and pixel-level sequence modeling suggests that powerful probabilistic models can serve as effective source coders when their predictive distributions are coupled with entropy coding \cite{deletang2024language,huang2024compression,li2025lmcompress,du2025llmlicvp,chen2025p2llm}.
This trend motivates the investigation of lossless pixel-level transmission as a complementary direction beyond semantic reconstruction.
Correspondingly, a separate source-channel coding (SSCC) architecture provides a natural and principled solution, particularly when combined with modern neural source models and reliability-oriented channel decoding techniques \cite{ren2025sscc,choukroun2022ecct,choukroun2024foundation,nguyen2024u}.
Under noisy channels, however, the source representation should be evaluated not only by compression efficiency, but also by how well it supports reliable exact recovery after imperfect channel decoding.

\subsection{Related Works}

Existing studies on semantic communications have mainly focused on task-oriented or distortion-oriented signal reconstruction.
Deep JSCC directly maps source signals to channel symbols and has shown strong robustness for wireless image transmission \cite{bourtsoulatze2019deep}.
Subsequent works further improve semantic visual transmission by introducing sparse transmission, adaptive learning, generative priors, and foundation models \cite{tong2025sparsesbc,lu2024self,jiang2024semantic,jiang2024large,liang2024generative,jiang2024multi}.
More recently, diffusion priors have also been introduced into semantic communication systems.
For example, DiT-JSCC employs generative diffusion modeling to improve perceptual reconstruction quality under bandwidth- or SNR-limited channels \cite{tan2026ditjscc}.
However, these diffusion-aided semantic communication methods are still designed for semantic or perceptual fidelity rather than bit-exact pixel recovery.
Therefore, digital or hybrid solutions based on SSCC remain indispensable when lossless reconstruction is required.
In such systems, source coding determines the number of bits to be protected, while channel coding determines whether these bits can be reliably delivered over a noisy link.
Classical channel codes, such as LDPC and Polar codes, provide strong digital protection mechanisms \cite{gallager1962low,arikan2009channel}.
Recent neural decoders further enhance reliability-oriented channel decoding \cite{nachmani2018deep,choukroun2022ecct,choukroun2024foundation,nguyen2024u}.
In this work, we follow the SSCC principle and focus on improving the source-coding component for lossless pixel-level image transmission.

From the source-coding viewpoint, lossless image transmission depends on accurate probability estimation before entropy coding.
Classical lossless image codecs improve compression by exploiting local context and adaptive prediction \cite{wu1997calic,weinberger2000loco,alakuijala2019jpegxl}.
Learned lossless codecs further strengthen conditional modeling with neural networks, flows, hierarchical latent variables, or bit-plane representations \cite{mentzer2019l3c,hoogeboom2019idf,kang2022pilc,zhang2024aribbps,li2025callic,bai2024dlpr,zhang2025fnlic}.
Meanwhile, arithmetic coding provides a principled interface between predictive distributions and binary lossless representations \cite{rissanen1976generalized,witten1987arithmetic,howard1994arithmetic}.
Recent large-model approaches further show that powerful sequence models can serve as effective compressors when their token probabilities are coupled with arithmetic coding \cite{deletang2024language,huang2024compression,valmeekam2023llmzip,du2025llmlicvp,li2025lmcompress}.
Representative image-oriented sequence models and coders, including iGPT and P$^{2}$-LLM, confirm that autoregressive next-token modeling can be adapted to pixel-level compression \cite{chen2020igpt,chen2025p2llm}.
Nevertheless, strict autoregression still suffers from a fixed coding order and inherently sequential decoding over long pixel streams.
This limits scalability and offers little flexibility in balancing compression performance against inference latency.
These deficiencies motivate non-autoregressive sequence modeling mechanisms that can exploit broader contexts and support more flexible token restoration orders.

Discrete diffusion models provide an appealing alternative by restoring masked tokens under bidirectional attention \cite{austin2021d3pm,li2022diffusionlm,sahoo2024mdlm,arriola2025block,nie2025llada}.
Concretely, diffusion models adopt a reverse process where coding starts from a fully masked state and progressively removes masks until completely recovering the clean pixel-token sequence.
Compared to predicting the next pixel in a rigid causal order, this non-causal mechanism is especially attractive for images with spatially heterogeneous predictability.
It also provides a progressively recoverable representation, since restored tokens can be reused as bidirectional context in later denoising steps.
Recent masked diffusion language models and block diffusion models further demonstrate that non-autoregressive or semi-autoregressive token restoration can provide flexible generation and decoding behaviors beyond strict left-to-right prediction \cite{sahoo2024mdlm,arriola2025block,nie2025llada,gong2025scaling}.
Consequently, diffusion models could alleviate the latency induced by strict one-token-per-step autoregression while offering a source representation better suited to noisy transmission.
However, existing diffusion priors in visual communication are mainly used for generative reconstruction or perceptual enhancement \cite{liang2024generative,tan2026ditjscc}, rather than for entropy-coded lossless compression and transmission.

Transforming a diffusion model into a lossless source coder is non-trivial.
Firstly, the deterministic left-to-right decoding order underlying autoregressive arithmetic coding no longer holds.
Therefore, to support arithmetic coding, the encoder and decoder must share the same denoising path together with aligned predictive probabilities at each denoising stage.
Secondly, during the iterative restoration in DDM~\cite{austin2021d3pm,sahoo2024mdlm,nie2025llada,arriola2025block}, confidence-based token selection is a common and effective sampling strategy~\cite{chang2022maskgit}.
However, 
confidence-driven restoration may repeatedly select locally smooth and highly correlated pixels \cite{witten1987arithmetic,howard1994arithmetic}, so the tokens encoded within the same jump step are predicted from the same pre-update context and may introduce redundant coding cost.
A similar clustering issue has also been observed in masked image generation, motivating low-discrepancy alternatives such as Halton scheduling~\cite{besnier2025halton}.
Furthermore, the step-wise denoising size is usually not matched to the reliability of the current context: early denoising states contain little information and require conservative updates, whereas late states can safely process more tokens.
Lastly, the predictive distributions of diffusion language models are optimized for denoising rather than directly for arithmetic coding, while the latter is sensitive to the exact probability table used for interval subdivision~\cite{witten1987arithmetic,howard1994arithmetic}.
Therefore, overconfident probabilities at highly masked states may increase the code length even when the top prediction is plausible.

\begin{table*}[!t]
\centering
\caption{Summary and comparison of related papers.}
\label{tab:comparison}
\renewcommand{\arraystretch}{1.25}
\begin{tabular*}{\textwidth}{
m{2.8cm}
>{\centering\arraybackslash}m{2.4cm}
>{\centering\arraybackslash}m{2.2cm}
>{\centering\arraybackslash}m{2.6cm}
m{6.0cm}}
\toprule
References 
& Lossless Pixel-Level Recovery 
& LLM-Based Source Coding 
& Bidirectional Diffusion Modeling 
& Brief Description \\
\hline

Bourtsoulatze \etal~\cite{bourtsoulatze2019deep},\newline Tong \etal~\cite{tong2025sparsesbc}
& $\EMPTYcircle$ 
& $\EMPTYcircle$ 
& $\EMPTYcircle$ 
& Performing distortion-oriented or task-oriented visual transmission with deep JSCC and semantic feature delivery. \\

Tan \etal~\cite{tan2026ditjscc},\newline Liang \etal~\cite{liang2024generative}
& $\EMPTYcircle$
& $\EMPTYcircle$
& $\CIRCLE$
& Employing diffusion or diffusion-transformer priors for semantic or perceptual JSCC reconstruction without bit-exact pixel recovery.  \\

Alakuijala \etal~\cite{alakuijala2019jpegxl},\newline Bai \etal~\cite{bai2024dlpr}
& $\CIRCLE$ 
& $\EMPTYcircle$ 
& $\EMPTYcircle$ 
& Improving lossless image compression through conventional codecs or learned residual probability modeling. \\

Ren \etal~\cite{ren2025sscc} 
& $\RIGHTcircle$ 
& $\CIRCLE$ 
& $\EMPTYcircle$ 
& Combining LLM-based lossless source decoding with reliable channel protection under the SSCC architecture. \\

Chen \etal~\cite{chen2025p2llm}
& $\CIRCLE$ 
& $\CIRCLE$ 
& $\EMPTYcircle$ 
& Utilizing autoregressive large language models for next-pixel probability estimation and arithmetic coding. \\

Chang \etal~\cite{chang2022maskgit},\newline Besnier \etal~\cite{besnier2025halton}
& $\EMPTYcircle$ 
& $\RIGHTcircle$ 
& $\CIRCLE$ 
& Enabling bidirectional masked token restoration with iterative generation and low-discrepancy scheduling. \\

\hline
This paper 
& $\CIRCLE$ 
& $\CIRCLE$ 
& $\CIRCLE$ 
& First applying discrete diffusion to lossless compression and integrating synchronized diffusion source coding into SSCC for exact pixel-level transmission. \\

\bottomrule
\multicolumn{5}{>{\vspace{-1mm}\footnotesize\itshape}r}{
Notations: \rm{${\CIRCLE}$} \emph{indicates fully included;} 
\rm{${\RIGHTcircle}$} \emph{means partially included;} 
\rm{${\EMPTYcircle}$} \emph{denotes not included.}
}

\end{tabular*}
\end{table*}

\subsection{Contributions}

In light of the above observations, this paper develops a Discrete Diffusion Model-based Separate Source-Channel Coding (DDM-SSCC) framework for pixel-level lossless image transmission. 
The proposed source codec turns diffusion denoising into a synchronized arithmetic-coding process and further improves it with schedule-aware denoising designs.
To tackle the practical mismatch between generation-oriented diffusion denoising and lossless arithmetic coding, we further develop the Halton-guided denoising rule, so as to improve the spatial coverage of denoised tokens. 
Besides, we introduce a cosine denoising schedule and mask-ratio-aware calibration module, thus adapting the denoising pace to context reliability and calibrating the probability tables consumed by arithmetic coding.
While providing a summary of the key differences between our algorithm and relevant literature in Table~\ref{tab:comparison}, we summarize the main contributions as follows:

\begin{itemize}
    \item We propose a synchronized discrete-diffusion source coding protocol for lossless pixel-token compression.
    Starting from a fully masked sequence, the encoder and decoder follow the same reverse denoising trajectory, use identical target positions and predictive distributions, and therefore make non-autoregressive diffusion compatible with arithmetic coding.
    To the best of our knowledge, this work is the first to make a DDM compatible with arithmetic coding and integrate it into SSCC for exact pixel-level image transmission.

    \item We analyze the denoising-position strategy and introduce a Halton-guided low-discrepancy denoising order that improves spatial coverage while remaining exactly reproducible at the decoder. Meanwhile, we design two mask-ratio-aware improvements for the diffusion source coder: a cosine denoising schedule that controls the number of processed tokens at each step, and a lightweight temperature calibration module that adapts the probability distribution used by arithmetic coding.

    \item We embed the proposed source coder into an SSCC transmission system with a fixed digital channel-protection backend and evaluate it under AWGN and Rayleigh fading channels.
    Experimental results show that the proposed source representation yields a more favorable exact-recovery operating point, while ablation studies verify the effectiveness of the proposed denoising order, schedule, and calibration modules.
\end{itemize}

\subsection{Paper Structure}

The remainder of this paper is organized as follows.
Section~\ref{sec:preliminaries_system_model} introduces the preliminaries, system model, and problem formulation.
Section~\ref{sec:proposed} presents the proposed discrete diffusion-based lossless source coding framework.
Section~\ref{sec:exp} reports the simulation settings and experimental results.
Section~\ref{sec:conclusion} concludes this paper.

\section{Preliminaries \& System Model}
\label{sec:preliminaries_system_model}
Beforehand, mainly used notations are summarized in Table~\ref{tab:key_notations}.

\begin{table}[!t]
\centering
\caption{List of key notations used in this paper.}
\label{tab:key_notations}
\renewcommand{\arraystretch}{1.15}
\begin{tabular*}{\columnwidth}{@{\extracolsep{\fill}}>{\centering\arraybackslash}p{0.2\columnwidth}p{0.75\columnwidth}@{}}
\toprule
\textbf{Notation} & \textbf{Definition} \\
\midrule
$\mathbf{X},\hat{\mathbf{X}}$ & Source image and reconstructed image. \\
$H,W,C$ & Numbers of image or patch rows, columns, and color channels. \\
$\mathbf{X}^{(p)},\mathbf{s}^{(p)}$ & The $p$-th image patch and its flattened pixel-value sequence. \\
$N$ & Number of maskable image tokens in one patch-level coding unit. \\
$\Phi_p(\cdot),\Phi_p^{-1}(\cdot)$ & Patch flattening and inverse patch reconstruction operations. \\
$\mathcal{T}(\cdot),\mathcal{T}^{-1}(\cdot)$ & One-to-one pixel-to-token mapping and inverse token-to-pixel mapping. \\
$\mathcal{V},\mathcal{D}$ & Full tokenizer vocabulary and valid $256$-symbol pixel-token dictionary. \\
$\mathbf{x}$ & Patch-level token sequence, where $x_0=\texttt{<bos>}$ is fixed and $x_i\in\mathcal{D}$ for $i\ge 1$. \\
$m$ & Absorbing mask token. \\
$\rho,\tilde{\rho}$ & True source distribution and model-based predictive distribution used for arithmetic coding. \\
$\mathbf{m},B(\mathbf{x}),R_s$ & Source bitstream, its length, and source rate in bits per pixel. \\
$C_s(\cdot),C_s^{-1}(\cdot)$ & Source encoder and decoder. \\
$C_e(\cdot),C_e^{-1}(\cdot)$ & Channel encoder/modulator and channel decoder. \\
$J,K,N_c,R_c$ & Number of channel blocks, information bits, codeword length, and code rate. \\
$h,\mathbf{n},\sigma_n^2$ & Channel coefficient, additive noise vector, and noise variance. \\
$\mathbf{y},\hat{\mathbf{m}}$ & Received signal and decoded source bitstream. \\
$\theta,f_{\theta}(\cdot)$ & Parameters and neural backbone of the diffusion source coder. \\
$T,t$ & Number of diffusion steps and step index. \\
$\mathbf{x}^{(t)}$ & Partially denoised token state before reverse step $t$. \\
$\mathcal{M}^{(t)},\mathcal{C}^{(t)}$ & Masked and visible token-position sets at step $t$. \\
$\beta_t,\bar{\alpha}_t$ & Forward masking probability and cumulative survival probability in the absorbing diffusion process. \\
$\mathbf{Z}^{(t)},z_{i,d}^{(t)}$ & Shifted pixel-token logits and the logit of token value $d$ at position $i$ during reverse step $t$. \\
$\hat{\rho}_{\theta}^{(t)}(d,i)$ & Model-predicted clean-token probability at position $i$. \\
$p_i^{(t)}(d)$ & Probability of token $d$ at position $i$ used for coding. \\
$\pi^{(t)},k_t$ & Ordered denoising list and number of tokens processed at reverse step $t$. \\
$r_i$ & Halton-based priority score assigned to position $i$. \\
$\varepsilon_t,\varepsilon_{\min},\varepsilon_{\max},\gamma$ & Mask-ratio-aware calibration temperature and its shared hyperparameters. \\
$\mathbf{u}^{(t)}$ & Synchronized encoder/decoder runtime state used in Algorithm~\ref{alg:synchronized_diffusion_codec}. \\
$\tau,\mu_{\tau}$ & Sampled forward diffusion step and corresponding training mask ratio. \\
$\tilde{\mathbf{x}}^{(\tau)},\mathcal{L}_{\mathrm{ft}}$ & Corrupted training sequence and fine-tuning loss. \\
$s_t,\alpha_t$ & Reverse progress variable and target cumulative denoising ratio in the cosine schedule. \\
$\mathcal{B}_{\mathrm{comm}},B_0$ & Communication-resource consumption and prescribed communication budget. \\
$P_{\mathrm{img}},\eta$ & Exact image-recovery probability and communication variables. \\
\bottomrule
\end{tabular*}
\end{table}

\subsection{Preliminaries}
\subsubsection{Lossless Source Coding with Probabilistic Sequence Models}

Let $\mathbf{x}_{1:N}=(x_1,\cdots,x_N)\in\mathcal{D}^N$ denote an $N$-length discrete source sequence with true distribution $\rho$.
Under ideal entropy coding, the code length is determined by the negative log-likelihood~\cite{cover2006elements},
namely $-\log_2 \rho(\mathbf{x}_{1:N})$.
Thus, a probabilistic model becomes a compressor by assigning high probability to the observed sequence.
For an autoregressive model,
\begin{equation}
\rho(\mathbf{x}_{1:N})
=
\prod_{i=1}^{N}
\rho(x_i\mid \mathbf{x}_{<i}),
\end{equation}
and the corresponding ideal codelength becomes
\begin{equation}
-\log_2 \rho(\mathbf{x}_{1:N})
=
-\sum_{i=1}^{N}
\log_2 \rho(x_i\mid \mathbf{x}_{<i}).
\label{eq:ar_codelength}
\end{equation}

In practice, the true distribution $\rho$ is approximated by a predictive model $\tilde{\rho}$, whose output probabilities can be coupled with arithmetic coding to produce a lossless bitstream $\mathbf{m}$~\cite{rissanen1976generalized,witten1987arithmetic,howard1994arithmetic}.
At coding step $i$, let $y_i\in\mathcal{D}$ be the target symbol and $\mathbf{c}_i$ be the shared encoder-decoder context.
Assume that the values in a dictionary $\mathcal{D}$ are arranged according to a fixed order shared by the encoder and decoder.
For a symbol value $d\in\mathcal{D}$, define the cumulative probability preceding $d$ as
$
F_i(d)=
\sum_{d'\prec d}\tilde{\rho}(d'\mid\mathbf{c}_i)
$. 
Starting from $\mathbb{I}_0=[L_0,U_0)=[0,1)$, encoding $y_i$ updates the interval as
\begin{equation}
\begin{aligned}
L_i
&=
L_{i-1}
+
\left(U_{i-1}-L_{i-1}\right)F_i(y_i),\\
U_i
&=
L_{i-1}
+
\left(U_{i-1}-L_{i-1}\right)
\left[F_i(y_i)+\tilde{\rho}(y_i\mid\mathbf{c}_i)\right].
\end{aligned}
\label{eq:generic_ac_interval}
\end{equation}
Along with encoding all symbols, the bitstream is chosen as a finite binary fraction in the final interval.
The decoder uses the same contexts and probability tables to identify the unique $i$-th symbol whose subinterval contains this fraction $\left[L_i, U_i \right)$, so losslessness requires \emph{synchronized} symbol order, probability tables, and context updates.

Finally, the resulting model-based codelength is approximately
\begin{equation}
\ell(\mathbf{x}_{1:n})
\simeq
-\sum_i \log_2 \tilde{\rho}(y_i\mid\mathbf{c}_i),
\label{eq:generic_model_codelength}
\end{equation}
with expected codelength characterized by the cross-entropy
\begin{equation}
H(\rho,\tilde{\rho})
=
\mathbb{E}_{\mathbf{x}_{1:n}\sim\rho}
\left[
-\sum_i
\log_2 \tilde{\rho}(y_i\mid \mathbf{c}_{i})
\right].
\label{eq:cross_entropy_rate}
\end{equation}
Thus, better probability estimation directly improves lossless compression, while non-standard coding orders additionally require deterministic synchronization.

\subsubsection{Discrete Diffusion-Based Sequence Modeling}

Discrete diffusion models define a forward corruption process and a learned reverse denoising process over discrete sequences~\cite{austin2021d3pm,sahoo2024mdlm}.
Let $\mathbf{x}^{(0)}\in\mathcal{D}^{N}$ be a clean sequence.
For consistency with the source-coding formulation in Section~\ref{sec:proposed}, we use the superscript $(t)$ to index the diffusion step and the subscript $\ell$ to index the sequence position.
A categorical forward process can be written as
\begin{equation}
q(\mathbf{x}^{(t)}\mid \mathbf{x}^{(t-1)})
=
\prod_{\ell=1}^{N}
q(x_{\ell}^{(t)}\mid x_{\ell}^{(t-1)}).
\end{equation}

This work focuses on the absorbing-state formulation, where a special mask token $m$ is used as the absorbing state~\cite{austin2021d3pm,sahoo2024mdlm}.
At each forward diffusion step, an unmasked symbol is either preserved or replaced by $m$:
\begin{equation}
q(x_{\ell}^{(t)}\mid x_{\ell}^{(t-1)})=
\begin{cases}
\beta_t, & x_{\ell}^{(t)}=m,\; x_{\ell}^{(t-1)}\neq m;\\
1-\beta_t, & x_{\ell}^{(t)}=x_{\ell}^{(t-1)},\; x_{\ell}^{(t-1)}\neq m;\\
1, & x_{\ell}^{(t)}=m,\; x_{\ell}^{(t-1)}=m,
\end{cases}
\label{eq:absorbing_forward}
\end{equation}
where $\beta_t$ is the masking probability.
Equivalently, conditioned on the clean sequence, the marginal corruption distribution is
\begin{equation}
q(x_{\ell}^{(t)}\mid x_{\ell}^{(0)})
=
\begin{cases}
\bar{\alpha}_t, & x_{\ell}^{(t)}=x_{\ell}^{(0)},\\
1-\bar{\alpha}_t, & x_{\ell}^{(t)}=m,\\
0, & \text{otherwise},
\end{cases}
\label{eq:absorbing_marginal}
\end{equation}
where $\bar{\alpha}_t=\prod_{\tau=1}^{t}(1-\beta_\tau)$.

The reverse process uses a neural model to estimate clean symbols from partially masked states.
For a state $\mathbf{x}^{(t)}$, the model outputs categorical distributions over the discrete alphabet.
Let $\mathcal{M}^{(t)}=\{\ell\in \{1, \cdots, N\} \mid x_{\ell}^{(t)}=m\}$ denote the masked position set until denoising step $t$ and let $\mathcal{C}^{(t)}=\{1,\cdots,N\}\setminus\mathcal{M}^{(t)}$ denote the visible position set, corresponding to a context $\mathbf{c}_t$.
The current partially observed state $\mathbf{x}^{(t)}$ therefore contains the denoised tokens at positions in $\mathcal{C}^{(t)}$ and mask tokens at positions in $\mathcal{M}^{(t)}$.

For any masked position $\ell\in\mathcal{M}^{(t)}$, the model provides a conditional distribution as
\begin{equation}
\hat{\rho}_{\theta}^{(t)}(d, \ell)
\triangleq 
p_{\theta}\!\left(x_{\ell}^{(0)}=d\mid \mathbf{x}^{(t)},\ell\right),
\quad d\in\mathcal{D}.
\label{eq:ddm_context_conditional}
\end{equation}
Here, conditioned on the whole partially observed state $\mathcal{C}^{(t)}$, $\hat{\rho}_{\theta}^{(t)}$ denotes the model-based estimate of the true source distribution $\rho$, and gives the probability table supplied to arithmetic coding.
Accordingly, a discrete diffusion model can be viewed as a probabilistic sequence model that supplies context-dependent symbol probabilities for masked positions.

\begin{figure*}[!tb]
    \centering
    \includegraphics[width=\textwidth]{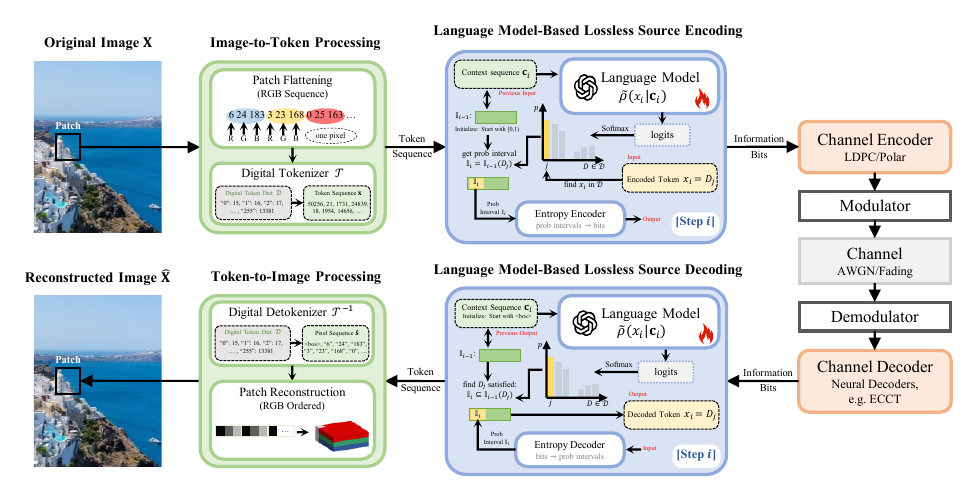}
    \vspace{-1.25cm}
    \caption{Overall DDM-SSCC pipeline for lossless pixel-level image transmission.}
    \label{fig:System_Model}
\end{figure*}

\subsection{System Model}
\label{subsec:model}

As illustrated in Fig.~\ref{fig:System_Model}, we consider pixel-level lossless image transmission over a noisy digital link under an SSCC architecture.
Consistent with patch-wise source processing in learned lossless image compression and large-model image coders \cite{chen2025p2llm}, the source coder operates patch by patch, and the final image bitstream is obtained by concatenating the compressed outputs of all patches. 
Without loss of generality, for a source image $\mathbf{X}\in\{0,1,\cdots,255\}^{H\times W\times C}$, where $H$, $W$, and $C$ denote the numbers of pixel rows, columns, and color channels, the image is first partitioned into non-overlapping patches.
The $p$-th patch is serialized into a pixel sequence
\begin{equation}
\mathbf{s}^{(p)}=\Phi_p(\mathbf{X}^{(p)})\in\{0,1,\cdots,255\}^{N_p},
\end{equation}
where $\Phi_p(\cdot)$ denotes patch flattening with RGB-ordered serialization and $N_p$ is the number of pixel values in the patch\footnote{For convenience of representation, we ignore the subscript $p$ in the subsequent description.}.
Each pixel value is then mapped to a unique token in a $256$-dimensional digital dictionary $\mathcal{D}$, and a beginning-of-sequence token \texttt{<bos>} is prepended:
\begin{equation}
\mathbf{x}=[x_0,\mathcal{T}(\mathbf{s})]\in\mathcal{V}^{N+1},
\end{equation}
where $\mathcal{V}$ denotes the tokenizer vocabulary, $x_0=\texttt{<bos>}$ is fixed and not entropy-coded, and $\mathcal{T}(\cdot)$ is a one-to-one pixel-to-token mapping from valid pixel values to the digital token dictionary $\mathcal{D}\subset\mathcal{V}$.
Only the image-token positions are entropy-coded, while the \texttt{<bos>} token is shared by the codec as a deterministic anchor. 
Therefore, lossless recovery in token space is equivalent to lossless recovery in the original image space.

By a compressor $C_s(\cdot)$, $\mathbf{x}$ is transformed into a binary bitstream $\mathbf{m}=C_s(\mathbf{x})$,
whose length is denoted by $B(\mathbf{x})=|\mathbf{m}|$.
The source rate of one coding unit is measured by
\begin{equation}
R_s=\frac{B(\mathbf{x})}{N} \ \bpp.
\end{equation}
where $\bpp$ stands for the number of bits per pixel.

For channel transmission, the compressed bitstream $\mathbf{m}$ is segmented into $J=\lceil B(\mathbf{x})/K\rceil$ information blocks, where $K$ denotes the number of information bits per channel-coding block. 
Each $K$-bit block is then encoded into an $N_c$-bit codeword by a channel code of rate $R_c=K/N_c$ and mapped to channel symbols through modulation.
The encoded source bitstream is then transmitted over a channel.
By denoting channel coding and modulation as $C_e(\cdot)$, the received signal can be written as
\begin{equation}
\mathbf{y}=h\,\widetilde{C}_e(\mathbf{m})+\mathbf{n},
\end{equation}
where $h$ is the channel coefficient and $\mathbf{n}\sim\mathcal{N}(\mathbf{0},\sigma_n^2\mathbf{I})$.
At the receiver, a channel decoder recovers an estimate $\hat{\mathbf{m}}$.

The source decoder then reconstructs the image as:
\begin{equation}
\hat{\mathbf{X}}=\Phi_p^{-1}\!\left(\mathcal{T}^{-1}\!\left(C_s^{-1}(\hat{\mathbf{m}})\right)\right).
\end{equation}

\subsection{Problem Formulation}

For pixel-level lossless transmission, it aims to accomplish exact end-to-end image recovery, namely
\begin{equation}
\hat{\mathbf{X}}=\mathbf{X}.
\end{equation}
Unlike distortion-oriented communication, any pixel mismatch is regarded as a failure.
Let $\theta$ denote the parameters of the diffusion-based source coder, and let $\eta$ collect the communication-side design variables such as channel rate and transmission energy.
The system-level objective can be written as
\begin{equation}
\begin{aligned}
\max_{\theta,\eta}\quad & P_{\text{img}}=\Pr(\hat{\mathbf{X}}=\mathbf{X})\\
\textrm{s.t.}\quad
& \mathbf{m}=C_s(\mathbf{x};\theta),\\
& \hat{\mathbf{m}}=C_e^{-1}(\mathbf{y};\eta),\\
& \mathcal{B}_{\text{comm}}(\eta,\mathbf{m})\le B_0,
\end{aligned}
\label{eq:problem}
\end{equation}
where $\mathcal{B}_{\text{comm}}$ denotes the prescribed communication budget.
In this paper, we focus on the source-coding component and treat the channel decoder as a fixed external module (e.g., error correction code transformer (ECCT)  \cite{choukroun2022ecct}). 
Even if the channel receiver is fixed, reducing the source burden and improving the stage-wise probability quality of arithmetic coding can still shift the full system toward a better reliability regime.
Moreover, the source representation itself affects how reliably exact recovery can be maintained under the same communication budget.
A diffusion-based source coder is therefore appealing not only for non-autoregressive modeling, but also for its progressive restoration process, where restored tokens can enhance later predictions as bidirectional context.
Consequently, the central question becomes whether a diffusion-based source coder can achieve competitive compression while yielding a more favorable exact-recovery operating point under the same communication budget.

\section{Proposed Discrete Diffusion-based Lossless Source Coding Framework}
\label{sec:proposed}

\subsection{Discrete Diffusion Model-based Source Codec with a Synchronized denoising schedule}
\label{subsec:ddm_source_codec}

As introduced in Section~\ref{sec:preliminaries_system_model}, the source image is first partitioned into patches, flattened into RGB-ordered pixel sequences, and converted into a digital token sequence by the one-to-one tokenizer $\mathcal{T}(\cdot)$. 
For a patch-level coding unit
$
\mathbf{x}=[x_0,x_1,\cdots,x_N]
$, instead of processing tokens in the raster-scan order of autoregressive coding, the discrete diffusion source coder starts from a fully masked state
\begin{equation}
\mathbf{x}^{(T)}=[x_0,m,\cdots,m],
\label{eq:ddm_init_state}
\end{equation}
where $m$ denotes the absorbing mask token, and progressively restores the masked image tokens until $\hat{\mathbf{x}}= \mathbf{x}^{(0)}$.
At reverse denoising step $t$, the current state $\mathbf{x}^{(t)}$ contains both restored and masked positions.
The diffusion backbone $f_{\theta}(\cdot)$ takes this partially denoised sequence as input and produces shifted pixel-token logits:
\begin{equation}
\mathbf{Z}^{(t)}
=
\operatorname{Shift}\!\left(f_{\theta}(\mathbf{x}^{(t)})\right)_{\mathcal{D}}.
\label{eq:shifted_reverse_logits}
\end{equation}
The shift operation follows the DiffuGPT adaptation recipe from autoregressive language models and is shared by fine-tuning and inference \cite{gong2025scaling}.

\begin{figure}[tb]
    \centering
    \includegraphics[width=\columnwidth]{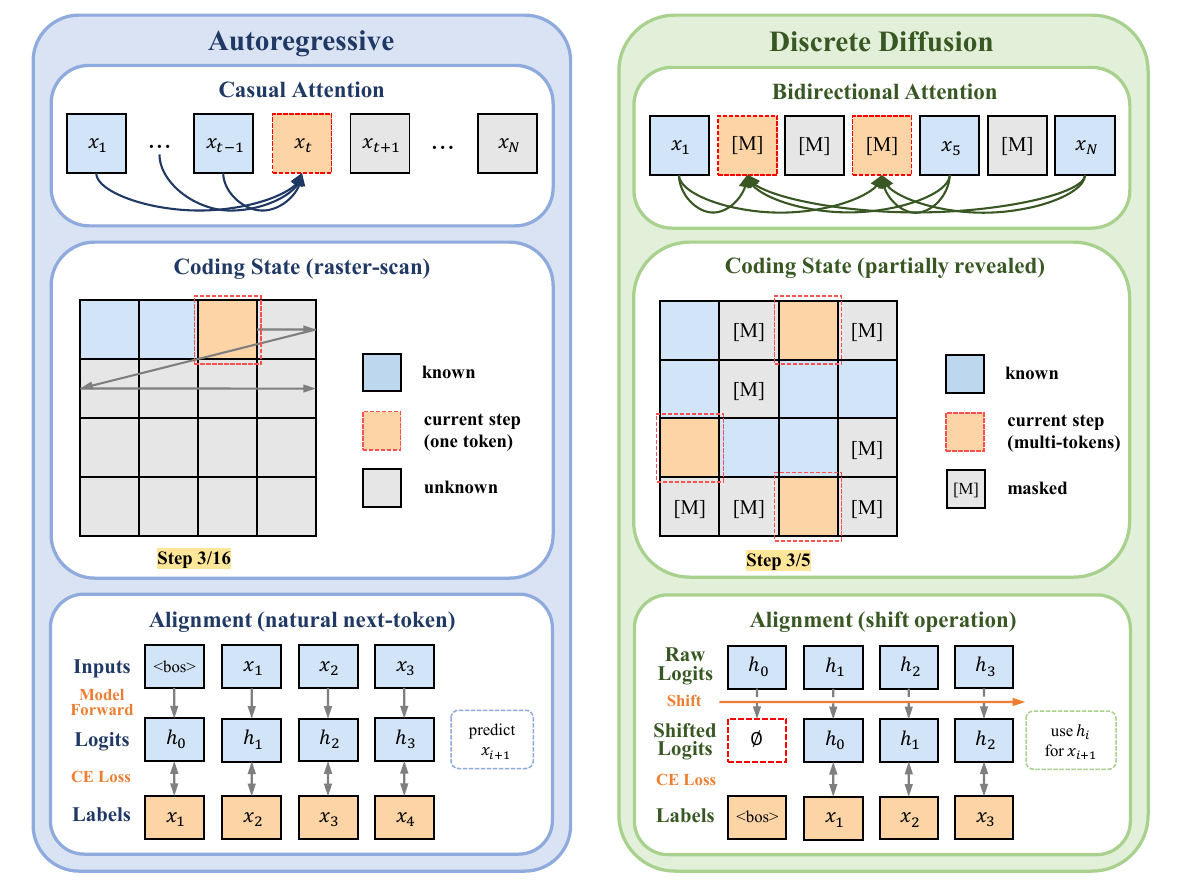}
    \vspace{-.5cm}
    \caption{Comparison between raster-order autoregressive coding and discrete diffusion restoration.}
    \label{fig:ar_vs_diffusion}
\end{figure}
Compared to autoregressive source coding, wherein the next coding context and the next probability table are naturally synchronized due to the left-to-right factorization, i.e., 
$p(\mathbf{x})=\prod_{i=1}^{N}p(x_i\mid x_0,x_1,\cdots,x_{i-1})$,  a discrete diffusion model predicts masked tokens from a shared partially denoised state under bidirectional attention~\cite{chang2022maskgit,sahoo2024mdlm}.
Thus, as illustrated in Fig.~\ref{fig:ar_vs_diffusion}, the coding context for position $i$ is the pair $(\mathbf{x}^{(t)},i)$ rather than a causal prefix, requiring an explicit denoising-path synchronization rule. In other words, compatibility with arithmetic coding requires the encoder and decoder to share the coding state, denoising positions, and probability distributions at each denoising step.
Specifically, as elaborated later in Section~\ref{subsec:halton_guided_denoising}, a synchronized scheduler selects an ordered denoising list
\begin{equation}
\pi^{(t)}=(i_1^{(t)},i_2^{(t)},\cdots,i_{k_t}^{(t)}),
\qquad
\pi^{(t)}\subseteq \mathcal{M}^{(t)},
\label{eq:denoising_list}
\end{equation}
where $k_t$ is the number of tokens processed at this denoising step (i.e., decoded tokens).
Since $\pi^{(t)}$ is deterministically generated from a shared scheduling rule and the current masked set, no side information about the denoising order is required.

Given the ordered denoising list $\pi^{(t)}$ and the shifted logits $\mathbf{Z}^{(t)} = [z_{i,d}^{(t)}] \in \mathbb{R}^{N\times \vert \mathcal{D}\vert }$ in \eqref{eq:shifted_reverse_logits},
for the $j$-th selected position $i_j^{(t)}$, Eq. \eqref{eq:ddm_context_conditional} can be formally written as 
\begin{equation}
p_j^{(t)}(d)
=
\hat{\rho}_{\theta,i_j^{(t)}}^{(t)}(d)
=
\frac{\exp(z_{i_j^{(t)},d}^{(t)})}
{\sum_{d'\in\mathcal{D}}\exp(z_{i_j^{(t)},d'}^{(t)})},
\quad d\in\mathcal{D}.
\label{eq:ddm_step_probability_table}
\end{equation}
The resulting discrete distribution is then used by arithmetic coding for interval subdivision.
Following \eqref{eq:generic_ac_interval}, the encoder narrows the arithmetic-coding interval using $\{p_j^{(t)}\}_{j=1}^{k_t}$, then fills the selected $k_t$ positions to obtain $\{x_{i_j^{(t)}}\}_{j=1}^{k_t}$ and then $\mathbf{x}^{(t-1)}$. Eventually, a reconstructed image $\hat{\mathbf{X}}$ can be obtained through inverse tokenization and patch reconstruction. Notably, the number of diffusion steps $T$ and the step-wise denoising size $k_t$ jointly control the trade-off between context refresh frequency and coding complexity.
\begin{figure*}[!tb]
    \centering
    \includegraphics[width=0.98\textwidth]{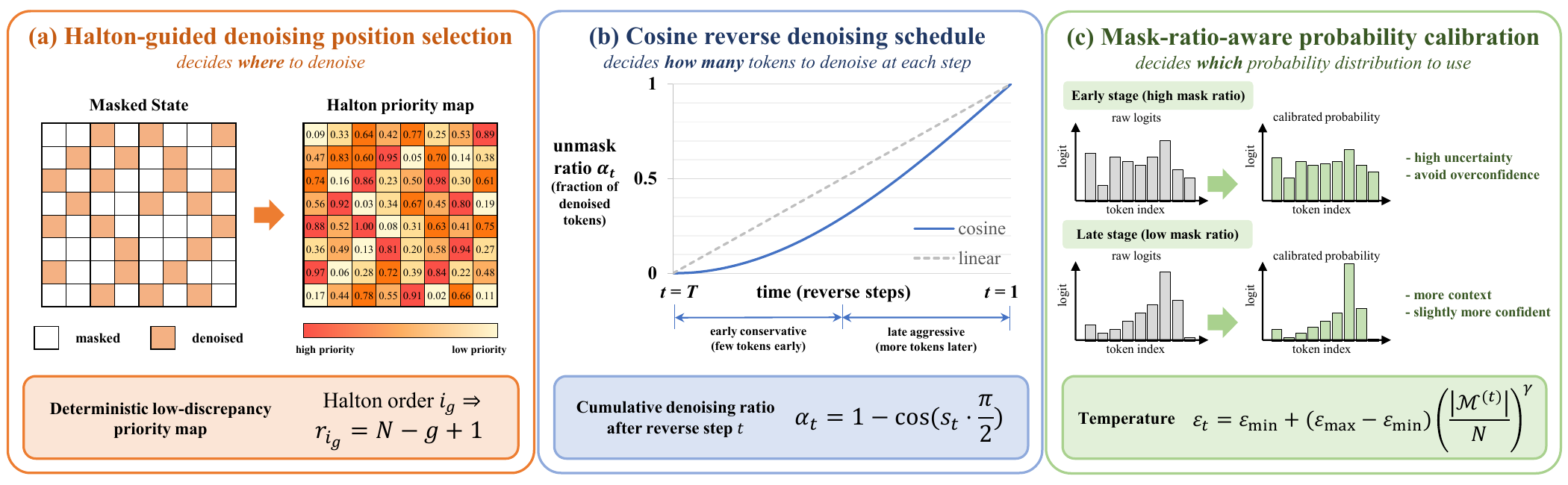}
    \vspace{-.5cm}
    \caption{Overview of the improved denoising strategy:
    (a) Halton-guided denoising position selection;
    (b) Cosine reverse denoising schedule;
    (c) Mask-ratio-aware probability calibration.}
    \vspace{-.5cm}
    \label{fig:improved_denoising_strategy}
\end{figure*}

\subsection{Halton-Guided Denoising Position Selection}
\label{subsec:halton_guided_denoising}

Generally, the synchronized codec can in principle use any deterministic denoising order, such as the confidence-based method in MaskGIT \cite{chang2022maskgit}.
However, as discussed in masked image generation~\cite{besnier2025halton}, a well-spread denoising set $\pi(t)$ can reduce same-step redundancy.
Therefore, we introduce Halton scheduling to guide denoising-position selection for improved spatial coverage.
For a patch with spatial size $H\times W$ and channel dimension $C$, let $N=HWC$ be the number of maskable image tokens.
As shown in Fig.~\ref{fig:improved_denoising_strategy}(a), each position $i\in\{1,\cdots,N\}$ is assigned according to a Halton-based priority map constructed by quantizing a Halton sequence to the patch grid and recording the first-visit order of discrete token positions. For lossless diffusion coding, the Halton priority map depends only on the patch geometry and shared deterministic rules, not on the source content, the logits, or any decoder-unknown information. 
Hence, this property also reduces same-step dependence while remaining exactly reproducible at the decoder.

Specifically, the Halton-based priority map is built from the radical-inverse sequence \cite{halton1964algorithm}.
For an integer $g\geq 1$, let
$g=\sum_{\ell=0}^{L}a_{\ell}b^{\ell}$
be its base-$b$ expansion, where $L$ is a non-negative integer and 
$a_{\ell}\in\{0,1,\cdots,b-1\}$.
The radical-inverse function is defined in terms of $a$ and $b$ as
\begin{equation}
\phi_b(g)
=
\sum_{\ell=0}^{L}a_{\ell}b^{-(\ell+1)}.
\label{eq:radical_inverse}
\end{equation}
That is, $\phi_b(g)$ maps the base-$b$ digits of $g$ to the fractional part in reverse order.
Given a set of pairwise coprime bases $\mathbf{b}=(b_1,\cdots,b_s)$, the $g$-th Halton point is obtained by
\begin{equation}
\mathbf{u}_g
=
\left[
\phi_{b_1}(g),\cdots,\phi_{b_s}(g)
\right]\in[0,1)^s.
\label{eq:halton_point}
\end{equation}
The points $\{\mathbf{u}_g\}_{g=1}^{\infty}$ form a Halton sequence in $[0,1)^s$.

In our patch-level codec, we use $\mathbf{b}=(2,3)$ for grayscale patches and $\mathbf{b}=(2,3,5)$ for RGB patches.
For an RGB patch, i.e., $C=3$, a Halton point $\mathbf{u}_q=(u_{g,1},u_{g,2},u_{g,3})$ is quantized to a discrete patch coordinate by
\begin{equation}
h_g = 1+\left\lfloor H u_{g,1}\right\rfloor,\quad
w_g = 1+\left\lfloor W u_{g,2}\right\rfloor,\quad
c_g = 1+\left\lfloor C u_{g,3}\right\rfloor ,
\label{eq:halton_quantization}
\end{equation}
where $h_g\in\{1,\cdots,H\}$, $w_g\in\{1,\cdots,W\}$, and $c_g\in\{1,\cdots,C\}$.
The corresponding flattened token position is
\begin{equation}
i_g
=
\bigl((h_g-1)W+(w_g-1)\bigr)C+c_g,
\quad i_g\in\{1,\cdots,N\}.
\label{eq:halton_flatten_index}
\end{equation}
Because multiple Halton points may be quantized to the same discrete position, we scan $g=1,2,\cdots$ but keep only the first visit to each position. For the position $i_g$, which is first visited by the $g$-th retained Halton point, the priority is assigned as $r_{i_g}=N-g+1$.
Then a deterministic ordering of all maskable positions can be obtained, where earlier visited positions receive larger priorities.
At denoising step $t$, the denoising list is obtained by selecting the top-$k_t$ active positions:
\begin{equation}
\pi^{(t)}
=
\operatorname{TopK}_{i\in\mathcal{M}^{(t)}}
\left(r_i, k_t\right),
\label{eq:halton_topk}
\end{equation}
where $\pi^{(t)}=(i_1^{(t)},\cdots,i_{k_t}^{(t)})$.
In implementation, already denoised and non-maskable \texttt{<bos>} token
 are assigned $-\infty$ before sorting, so only positions in $\mathcal{M}^{(t)}$ can be selected.

\begin{figure}[tb]
    \centering
    \includegraphics[width=\columnwidth]{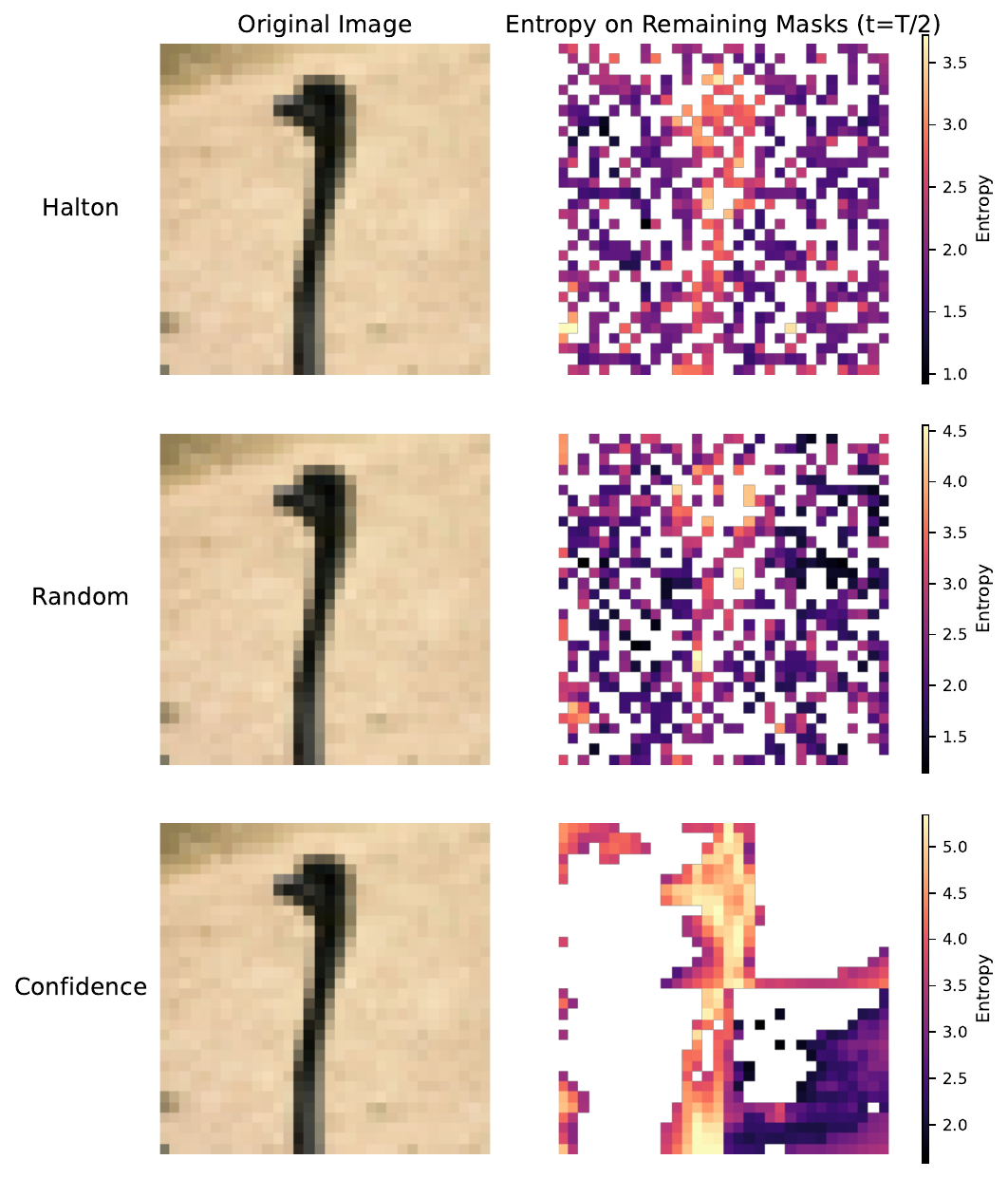}
    \captionsetup{skip=2pt} 
    \caption{Visualization of remaining-mask entropy at the middle denoising stage for different denoising-position selection rules.}
    \label{fig:ablation_halton_visual}
\end{figure}

Fig.~\ref{fig:ablation_halton_visual} visualizes the remaining-mask entropy at the middle denoising stage. 
Compared with confidence-based MaskGIT-style selection\cite{chang2022maskgit}, the Halton priority map is independent of the model logits and therefore avoids repeatedly selecting only the easiest local regions.
Compared with independent pseudo-random priorities, 
the Halton priority map provides a quasi-random but more evenly distributed ordering over the patch. 
In this sense, Halton-guided denoising is a low-discrepancy derandomization of randomized denoising: it keeps the non-confidence-driven property while producing a more uniformly covered denoising trajectory.


\subsection{Mask-Ratio-Aware Denoising Schedule}
\label{subsec:mask_ratio_schedule_calibration}

After determining \emph{where} to denoise by the Halton rule (i.e., $\pi^{(t)}$), the codec still needs to decide \emph{how many} tokens to process ($k_t$) and \emph{which} probability table to use.
We address these issues with a cosine schedule and mask-ratio-aware calibration, inspired by masked token generation and discrete diffusion models~\cite{chang2022maskgit,sahoo2024mdlm}.

\subsubsection{Cosine Schedule}
Recalling that $\mathcal{M}^{(t)}$ denotes the set of positions that remain masked before denoising step $t$, while $N$ is the number of maskable image tokens. A simple linear schedule $k_t^{\mathrm{lin}}=\frac{|\mathcal{M}^{(t)}|}{t}$ processes an approximately uniform fraction of the remaining masked tokens over the remaining denoising steps.
This rule guarantees completion at $t=1$, but it does not distinguish sparse early contexts from richer late contexts~\cite{chang2022maskgit}. We therefore resort to a mask-ratio-aware cosine schedule.
Without loss of generality, let
$s_t=\frac{T-t+1}{T}$, $\forall t=T,T-1,\cdots,1$, and the target cumulative denoising ratio after denoising step $t$ is defined as
\begin{equation}
\alpha_t
= \begin{cases}
1-\cos\left(
s_t\cdot\frac{\pi}{2}
\right), & t=T,T-1,\cdots,1\\
0, & t = T+1,
\end{cases}.
\label{eq:cosine_cumulative_denoising_ratio}
\end{equation}
where $\alpha_{T+1}=0$ for notational convenience.
Accordingly, the number of newly processed tokens at denoising step $t$ is
\begin{equation}
k_t
=
N \left(\alpha_t - \alpha_{t+1} \right),
\label{eq:cosine_step_size}
\end{equation}
except that the final step uses $k_1=|\mathcal{M}^{(1)}|$ to guarantee complete restoration.

As shown in Fig.~\ref{fig:improved_denoising_strategy}(b), this cosine schedule is conservative at the beginning because $\alpha_t$ grows slowly when $t$ is close to $T$.
Only a few anchor tokens are denoised when the available context is limited.
As $t$ decreases, more tokens have been restored and the context becomes richer; the schedule then increases the step size and processes more tokens per denoising step.
Thus, the cosine schedule improves early-stage probability reliability while retaining the jump-step efficiency of diffusion coding.

\subsubsection{Mask Ratio-Aware Calibration}

Since the accuracy of diffusion predictions changes markedly with the remaining mask ratio, using the raw logits at every denoising step is suboptimal for arithmetic coding, and temperature scaling provides a simple way to correct this kind of probability miscalibration \cite{guo2017calibration}.
We therefore re-write the distribution for the $j$-th denoising position $i_j^{(t)}$ in Eq. \eqref{eq:ddm_step_probability_table} as
\begin{equation}
p_j^{(t)}(d)
=
\frac{\exp(z_{i_j^{(t)},d}^{(t)}/\varepsilon_t)}
{\sum_{d'\in\mathcal{D}}\exp(z_{i_j^{(t)},d'}^{(t)}/\varepsilon_t)},
\label{eq:calibrated_prob}
\end{equation}
where the mask-ratio-aware temperature is defined as
\begin{equation}
\varepsilon_t
=
\varepsilon_{\min}+(\varepsilon_{\max}-\varepsilon_{\min})
\left(\frac{|\mathcal{M}^{(t)}|}{N}\right)^{\gamma},
\label{eq:temperature}
\end{equation}
where $\varepsilon_{\min}$, $\varepsilon_{\max}$, and $\gamma$ are shared hyperparameters.
As shown in Fig.~\ref{fig:improved_denoising_strategy}(c), when many tokens remain masked, $\varepsilon_t>1$ softens overconfident predictions and reduces high-probability assignments in arithmetic coding.
When only a few tokens remain masked, $\varepsilon_t<1$ sharpens the distribution and improves compression efficiency.

\begin{algorithm}[t]
\caption{Domain-Adapted Fine-Tuning of DiffuGPT}
\label{alg:domain_adapted_finetuning}
\begin{algorithmic}[1]
\REQUIRE Image mini-batches, pretrained DiffuGPT $f_{\theta}$, pixel-token dictionary $\mathcal{D}$, mask token $m$,  maximum forward step $T$.
\ENSURE Fine-tuned model $f_{\theta}$
\FOR{$r=1$ to epochs}
    \STATE Tokenize each image patch as $\mathbf{x}=[x_0,x_1,\cdots,x_N]$, where $x_0=\texttt{<bos>}$ and $x_i\in\mathcal{D}$ for $i\geq 1$.
    \STATE Sample a forward diffusion step $\tau\in\{1,\cdots,T\}$ and compute the corresponding mask ratio $\mu_{\tau}=1-\bar{\alpha}_{\tau}$.
    \STATE Form a corrupted sequence $\tilde{\mathbf{x}}^{(\tau)}$ by replacing each image-token position with $m$ independently with probability $\mu_{\tau}$; keep $x_0$ unchanged.
    \STATE Let $\mathcal{M}^{(\tau)}=\{\ell\in\{1,\cdots,N\}:\tilde{x}^{(\tau)}_\ell=m\}$ be the masked image-token set.
    \IF{$\mathcal{M}^{(\tau)}=\emptyset$}
        \STATE Continue.
    \ENDIF
    \STATE Compute shifted pixel-token logits $\mathbf{Z}^{(\tau)}\leftarrow\operatorname{Shift}(f_{\theta}(\tilde{\mathbf{x}}^{(\tau)}))_{\mathcal{D}}$ by Eq.~\eqref{eq:shifted_reverse_logits}.
    \STATE Compute $\mathcal{L}_{\mathrm{ft}}(\theta)$ by Eq.~\eqref{eq:train_loss} over $\mathcal{M}^{(\tau)}$ and update $\theta$.
\ENDFOR
\RETURN $f_{\theta}$
\end{algorithmic}
\end{algorithm}

\begin{algorithm}[t]
\caption{Synchronized Diffusion Arithmetic Coding and Decoding}
\label{alg:synchronized_diffusion_codec}
\begin{algorithmic}[1]
\REQUIRE Fine-tuned model $f_{\theta}$, pixel dictionary $\mathcal{D}$, mask token $m$, Halton map $\mathcal{H}$, steps $T$, calibration parameters $(\varepsilon_{\min},\varepsilon_{\max},\gamma)$, arithmetic coder
\REQUIRE Encoder input: clean sequence $\mathbf{x}=[x_0,x_1,\cdots,x_N]$; decoder input: bitstream
\ENSURE Encoder output: bitstream; decoder output: reconstructed sequence $\hat{\mathbf{x}}$
\STATE Initialize the synchronized state $\mathbf{u}^{(T)}\leftarrow[x_0,m,\cdots,m]$ and masked set $\mathcal{M}^{(T)}\leftarrow\{1,\cdots,N\}$.
\FOR{$t=T,T-1,\cdots,1$ \textbf{while} $\mathcal{M}^{(t)}\neq\emptyset$}
    \STATE Compute shifted pixel-token logits $\mathbf{Z}^{(t)}\leftarrow\operatorname{Shift}(f_{\theta}(\mathbf{u}^{(t)}))_{\mathcal{D}}$ by Eq.~\eqref{eq:shifted_reverse_logits}.
    \STATE Determine $k_t$ by Eq.~\eqref{eq:cosine_step_size}, clipped to $1\leq k_t\leq|\mathcal{M}^{(t)}|$, with $k_1=|\mathcal{M}^{(1)}|$.
    \STATE Select the ordered denoising list $\pi^{(t)}=\operatorname{TopK}_{\ell\in\mathcal{M}^{(t)}}(r_\ell,k_t)$ by Eq.~\eqref{eq:halton_topk}.
    \STATE Compute the remaining mask ratio $|\mathcal{M}^{(t)}|/N$ and the temperature $\varepsilon_t$ by  Eq.~\eqref{eq:temperature}.
    \FOR{each position $\ell\in\pi^{(t)}$}
        \STATE Form $p_\ell^{(t)}(d)=\operatorname{softmax}(z_{\ell,d}^{(t)}/\varepsilon_t)$ over $d\in\mathcal{D}$ by Eq.~\eqref{eq:calibrated_prob}.
        \STATE Compute the distribution function $F_\ell^{(t)}(d)=\sum_{d'\prec d}p_\ell^{(t)}(d')$ under the fixed ordering of $\mathcal{D}$.
        \IF{encoder mode}
            \STATE Arithmetic-encode the ground-truth token $x_i$ using
            $F_\ell^{(t)}(x_\ell)$ and $p_\ell^{(t)}(x_\ell)$ as in Eq.~\eqref{eq:generic_ac_interval},
            and set $u_\ell^{(t-1)}\leftarrow x_\ell$.
        \ELSE
            \STATE Arithmetic-decode $\hat{x}_\ell$ from the current interval using
            $F_\ell^{(t)}(\cdot)$ and $p_\ell^{(t)}(\cdot)$ as in Eq.~\eqref{eq:generic_ac_interval},
            and set $u_\ell^{(t-1)}\leftarrow \hat{x}_\ell$.
        \ENDIF
    \ENDFOR
    \STATE Copy all unprocessed positions from $\mathbf{u}^{(t)}$ to $\mathbf{u}^{(t-1)}$ and set $\mathcal{M}^{(t-1)}\leftarrow\mathcal{M}^{(t)}\setminus\pi^{(t)}$.
\ENDFOR
\STATE Output the terminated bitstream in encoder mode, or $\hat{\mathbf{x}}\leftarrow\mathbf{u}^{(0)}$ in decoder mode.
\end{algorithmic}
\end{algorithm}

\subsection{Fine-Tuning and Inference Procedures}
\label{subsec:training_inference_procedures}

\subsubsection{Domain-Adapted Fine-Tuning}
\label{subsubsec:domain_adapted_finetuning}
During fine-tuning, the pretrained DiffuGPT backbone is adapted to image-token denoising by masking pixel-token positions and keeping the \texttt{<bos>} token fixed, applying the same logit-shift operation that will be used during inference, and computing the loss only over the $256$ valid pixel tokens.
Since DiffuGPT is pretrained with a next-token objective, the hidden state at position $\ell-1$ predicts token $x_\ell$. 
Thus, for a corrupted sequence $\tilde{\mathbf{x}}^{(\tau)}$ generated at sampled forward step $\tau$, we shift the raw logits by one position and restrict the output vocabulary to the pixel-token dictionary $\mathcal{D}$ as in Eq. \eqref{eq:shifted_reverse_logits}.
The sampled step $\tau$ determines the training mask ratio
$
\mu_{\tau}=1-\bar{\alpha}_{\tau}$, 
which is consistent with the absorbing forward marginal in Eq.~\eqref{eq:absorbing_marginal}.
Only image-token positions are randomly replaced by the mask token $m$, while \texttt{<bos>} is kept unchanged.
Let $\mathcal{M}^{(\tau)}=\{\ell\in\{1,\cdots,N\} \mid \tilde{x}^{(\tau)}_\ell=m\}$ denote the masked image-token positions.
For each $\ell\in\mathcal{M}^{(\tau)}$, the training distribution $p_{\theta}^{(\tau)}(d\mid\tilde{\mathbf{x}}^{(\tau)},\ell)$ is given by Eq. \eqref{eq:calibrated_prob}.
The fine-tuning objective is then written as
\begin{equation}
\mathcal{L}_{\mathrm{ft}}(\theta)
=
\mathbb{E}_{\tau,\mathbf{x}}
\left[
\frac{w(\mu_{\tau})}{|\mathcal{M}^{(\tau)}|}
\sum_{\ell\in\mathcal{M}^{(\tau)}}
-\log p_{\theta}^{(\tau)}(x_\ell \mid\tilde{\mathbf{x}}^{(\tau)},\ell)
\right],
\label{eq:train_loss}
\end{equation}
where $w(\mu_{\tau})$ is the diffusion-rate weight.
Mini-batches with no valid masked pixel tokens are ignored.
The complete fine-tuning procedure is summarized in Algorithm~\ref{alg:domain_adapted_finetuning}.

\subsubsection{Synchronized Inference}
\label{subsubsec:synchronized_inference}

At inference time, compression and decompression use the same masked state $\mathbf{u}^{(t)}$,  masked set $\mathcal{M}^{(t)}$, denoising list $\pi^{(t)}$, schedule $k_t$, temperature $\varepsilon_t$, shifted logits $\mathbf{Z}^{(t)}$, and normalized probability tables $p_i^{(t)}$. 
For each selected position $i\in\pi^{(t)}$, the encoder arithmetically encodes the ground-truth pixel value $x_i\in\mathcal{D}$ according to $p_i^{(t)}(d)$, whereas the decoder decodes $\hat{x}_i\in\mathcal{D}$ from the bitstream using the same $p_i^{(t)}(d)$. 
Since $\pi^{(t)}$ is generated by the shared Halton-guided denoising rule in Eq.~\eqref{eq:denoising_list}, all shared parameters, including the  Halton priority map, denoising rule, denoising schedule, and calibration parameters, must be identical at both sides. Algorithm~\ref{alg:synchronized_diffusion_codec} summarizes the inference procedures.

\begin{table}[!t]
\centering
\caption{Key Parameter Settings for the Simulation and Experiments}
\label{tab:hyperparams}
\small
\renewcommand{\arraystretch}{1.15}
\begin{tabular*}{\columnwidth}{@{\extracolsep{\fill}}p{0.6\columnwidth}p{0.34\columnwidth}@{}}
\toprule
\textbf{Parameter Description} & \textbf{Value} \\
\midrule
\multicolumn{2}{l}{\textbf{Source Coder Training (DiffuGPT)}} \\
\midrule
Base architecture & GPT-2 \\
Training dataset & DIV2K-HR \\
Vocabulary size (original $\rightarrow$ valid pixel tokens) & $50{,}257 \rightarrow 256$ \\
Patch size (spatial resolution) & $16 \times 16$ \\
Learning rate & $3 \times 10^{-4}$ \\
Effective batch size & $256$ \\
Number of training epochs & $3$ \\

\midrule
\multicolumn{2}{l}{\textbf{Synchronized Diffusion Inference}} \\
\midrule
Inference diffusion steps $T$ & $20$ (default) \\
Arithmetic coder precision & $32$ bits \\
Temperature scaling bounds $[\varepsilon_{\min}, \varepsilon_{\max}]$ & $[0.9, 1.2]$ \\
Temperature scaling exponent $\gamma$ & $1.5$ \\

\midrule
\multicolumn{2}{l}{\textbf{Channel Protection (ECCT)}} \\
\midrule
Learning rate & $10^{-3}$ \\
Batch size & $256$ \\
Number of encoder layers & $6$ \\
Dimension of embedding & $128$ \\
Number of attention heads & $8$ \\
Coding rate & $1/2$ \\

\midrule
\multicolumn{2}{l}{\textbf{Evaluation Protocol}} \\
\midrule
Test datasets & CIFAR10, DIV2K-LR-X4, Kodak \\
Channel models & AWGN, Rayleigh \\
Quality metrics & PSNR, SSIM \\

\bottomrule
\end{tabular*}
\end{table}

\begin{figure*}[!t]
    \centering
    \includegraphics[width=\textwidth]{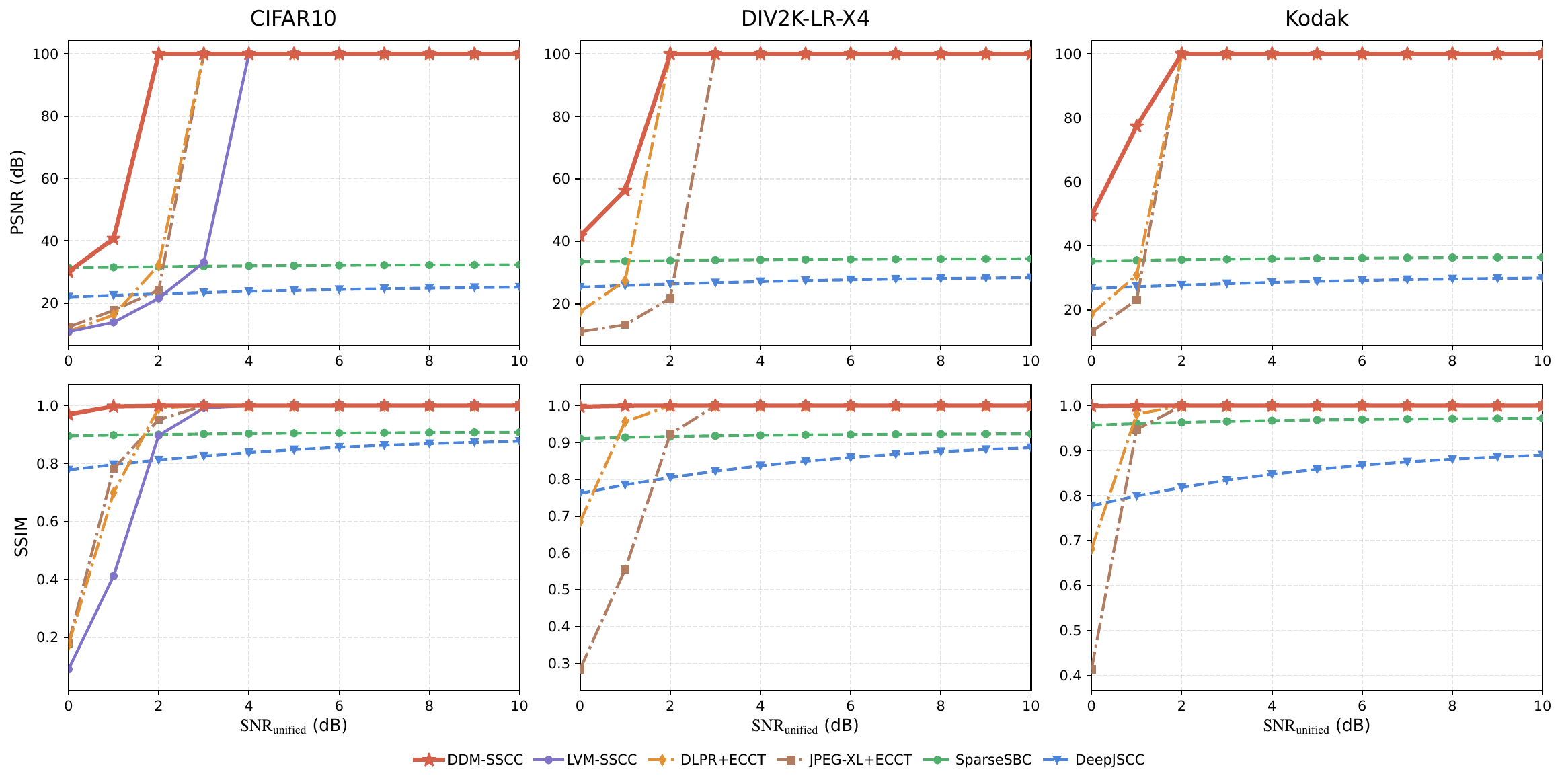}
    \caption{Main performance comparison on CIFAR10, DIV2K-LR-X4 and Kodak under AWGN channels. The upper row reports PSNR and the lower row reports SSIM. Due to the awfully high running-time, the result of LVM-SSCC for DIV2K-LR-X4 (validation) and Kodak can not be obtained at a reasonable amount time (several hours on NVIDIA GeForce RTX 4090 for a single image, consistent with the results in \cite{li2025lmcompress, chen2025p2llm}), and omitted here.}
    \vspace{-.5cm}
    \label{fig:main_AWGN}
\end{figure*}

\begin{figure*}[!tb]
    \centering
    \includegraphics[width=\textwidth]{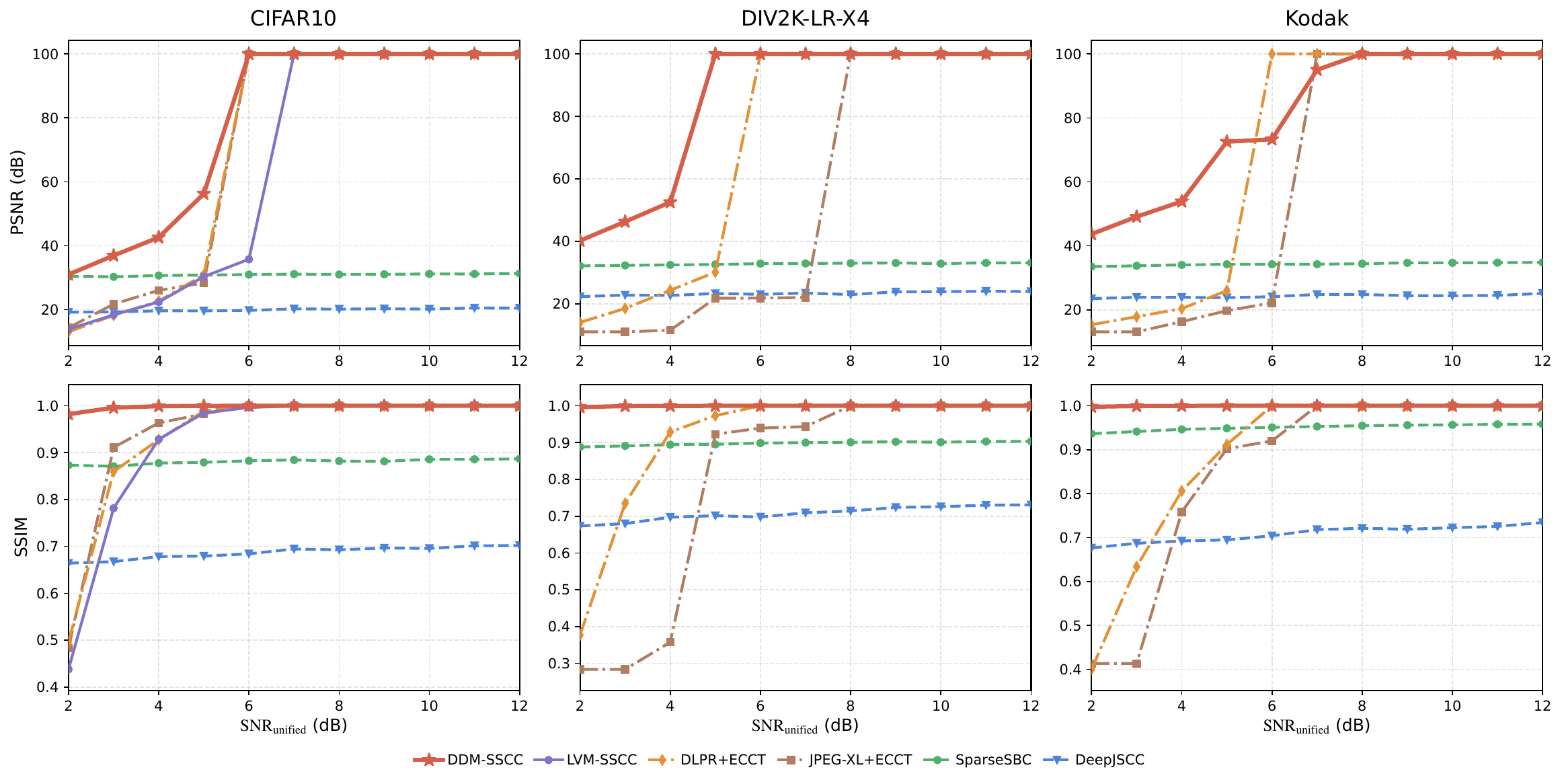}
    \caption{Main performance comparison on CIFAR10, DIV2K-LR-X4 and Kodak under Rayleigh fading channels. The upper row reports PSNR and the lower row reports SSIM. Due to the awfully high running-time, the result of LVM-SSCC for DIV2K-LR-X4 (validation) and Kodak can not be obtained at a reasonable amount time (several hours on NVIDIA GeForce RTX 4090 for a single image, consistent with the results in \cite{li2025lmcompress, chen2025p2llm}), and omitted here.}
    \vspace{-.5cm}
    \label{fig:main_Rayleigh}
\end{figure*}

\begin{figure*}[!tb]
    \centering
    \includegraphics[width=.875\textwidth]{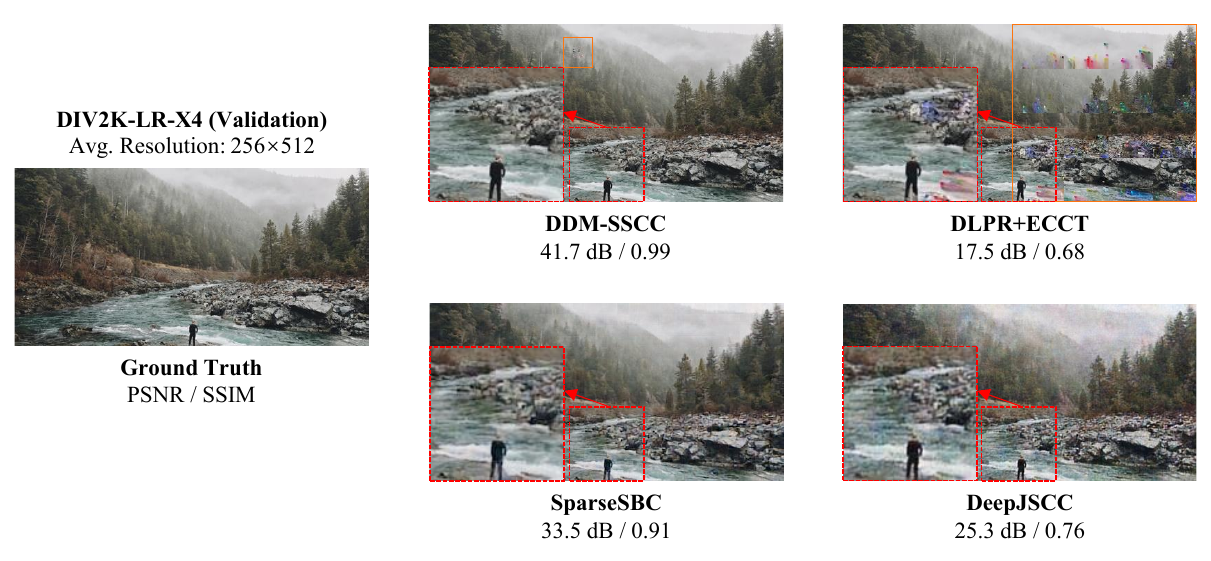}
    \captionsetup{skip=2pt} 
    \caption{Representative reconstruction example at $\text{SNR}_{\text{unified}}=0$ dB on DIV2K-LR-X4.}
    \vspace{-.5cm}
    \label{fig:visual_div2k}
\end{figure*}

\begin{figure*}[!tb]
    \centering
    \includegraphics[width=\textwidth]{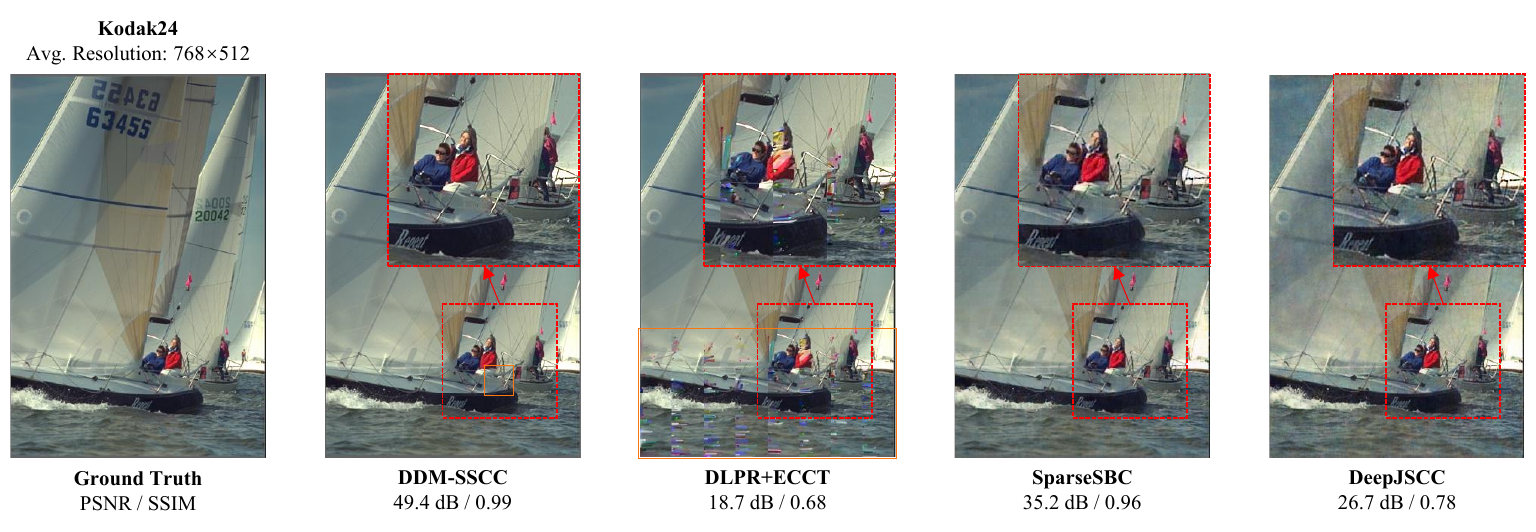}
    \captionsetup{skip=2pt} 
    \caption{Representative reconstruction example at $\text{SNR}_{\text{unified}}=0$ dB on Kodak.}
    \vspace{-.5cm}
    \label{fig:visual_Kodak}
\end{figure*}

\section{Experiments}
\label{sec:exp}

\subsection{Experimental Settings}

We evaluate the proposed DDM-SSCC on CIFAR10, DIV2K-LR-X4 (validation) and Kodak \cite{krizhevsky2009cifar,agustsson2017ntire,franzen1999kodak}, which respectively represent low-resolution and higher-resolution image transmission scenarios. 
For both datasets, the source coder follows the patch-wise processing in Section~\ref{subsec:model}, and the final image is obtained by concatenating all patch-level bitstreams and reassembling the decoded patches.

To focus on the source-coding effect, all digital schemes use the same channel-protection backend, namely a rate-$1/2$ channel code together with an ECCT-enhanced receiver \cite{choukroun2022ecct}.
All schemes are tested under both additive white Gaussian noise (AWGN) and Rayleigh channels. 
We primarily compare DDM-SSCC with four representative baselines: deep lossy plus residual coding (DLPR)+ECCT \cite{bai2024dlpr}, JPEG-XL+ECCT \cite{alakuijala2019jpegxl}, SparseSBC \cite{tong2025sparsesbc}, and DeepJSCC \cite{bourtsoulatze2019deep}. 
To highlight the difference between one-token causal decoding and the proposed multi-token reverse restoration, we further include an iGPT-based autoregressive source coder, denoted by LVM-SSCC \cite{chen2020igpt}. 
Mainly used hyperparameters are summarized in Table~\ref{tab:hyperparams}.

The reconstruction quality is measured by peak signal-to-noise ratio (PSNR) and structural similarity (SSIM). 
Since exact reconstruction gives zero MSE and hence an infinite PSNR under the standard definition, we report such cases as $\mathrm{PSNR}=100$ dB only for finite numerical display \cite{benazza2007block}.
To compare heterogeneous transmission schemes under the same communication budget, the horizontal axis is the unified signal-to-noise ratio (SNR)~\cite{cover2006elements}, denoted by $\text{SNR}_{\text{unified}}$. 
In other words, $\text{SNR}_{\text{unified}}$ aligns the same total communication budget.
Meanwhile, $\text{SNR}_{\text{unified}}$ serves as the experimental instantiation of the prescribed budget constraint $\mathcal{B}_{\text{comm}}(\eta,\mathbf{m})\le B_0$ in Eq.\eqref{eq:problem}: under a fixed total energy and reference channel-use budget, a source coder that produces a shorter protected bitstream can operate at a more favorable physical SNR.

\subsection{Experimental Results}
\subsubsection{Main Results}
Figs.~\ref{fig:main_AWGN} and~\ref{fig:main_Rayleigh} compare the end-to-end transmission performance of different schemes. 
Overall, DDM-SSCC achieves the most favorable performance. 
On three datasets under AWGN, DDM-SSCC reaches perfect reconstruction when $\text{SNR}_{\text{unified}}$ is larger than $2$ dB, a smaller threshold than the requirement for DLPR+ECCT and JPEG-XL+ECCT. 
Comparatively, the semantic communication baselines show a different behavior and remain far from the exact-recovery region. 
In other words, compared with semantic-oriented transmission, DDM-SSCC is much better aligned with fidelity-sensitive pixel-level delivery. 
A similar phenomenon is also observed for the Rayleigh fading channel. 
The advantage becomes even more pronounced on DIV2K-LR-X4 and Kodak, where high-resolution textures impose stronger requirements on probability modeling.
Notably, compared with the iGPT-based autoregressive source coder LVM-SSCC
, the proposed DDM-SSCC consistently moves the recovery threshold to a lower SNR region, validating the advantage of bidirectional masked restoration over one-token causal decoding for pixel-level lossless transmission.

\begin{figure}[!tb]
    \centering
    \includegraphics[width=.825\columnwidth]{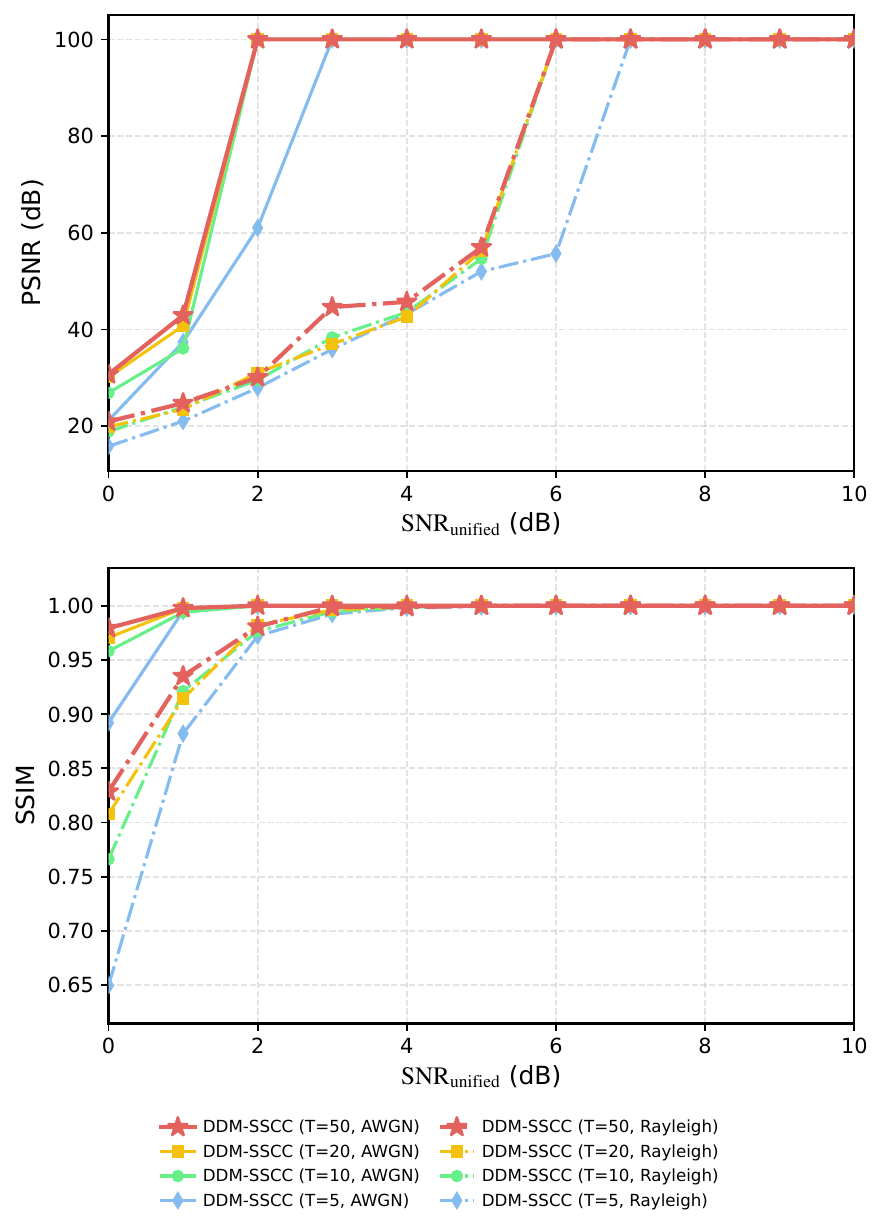}
    \captionsetup{skip=2pt} 
    \caption{Impact of the diffusion step number on CIFAR10.}
    \label{fig:diffu_step_comparison}
\end{figure}

\begin{figure}[!tb]
    \centering
    \includegraphics[width=.875\columnwidth]{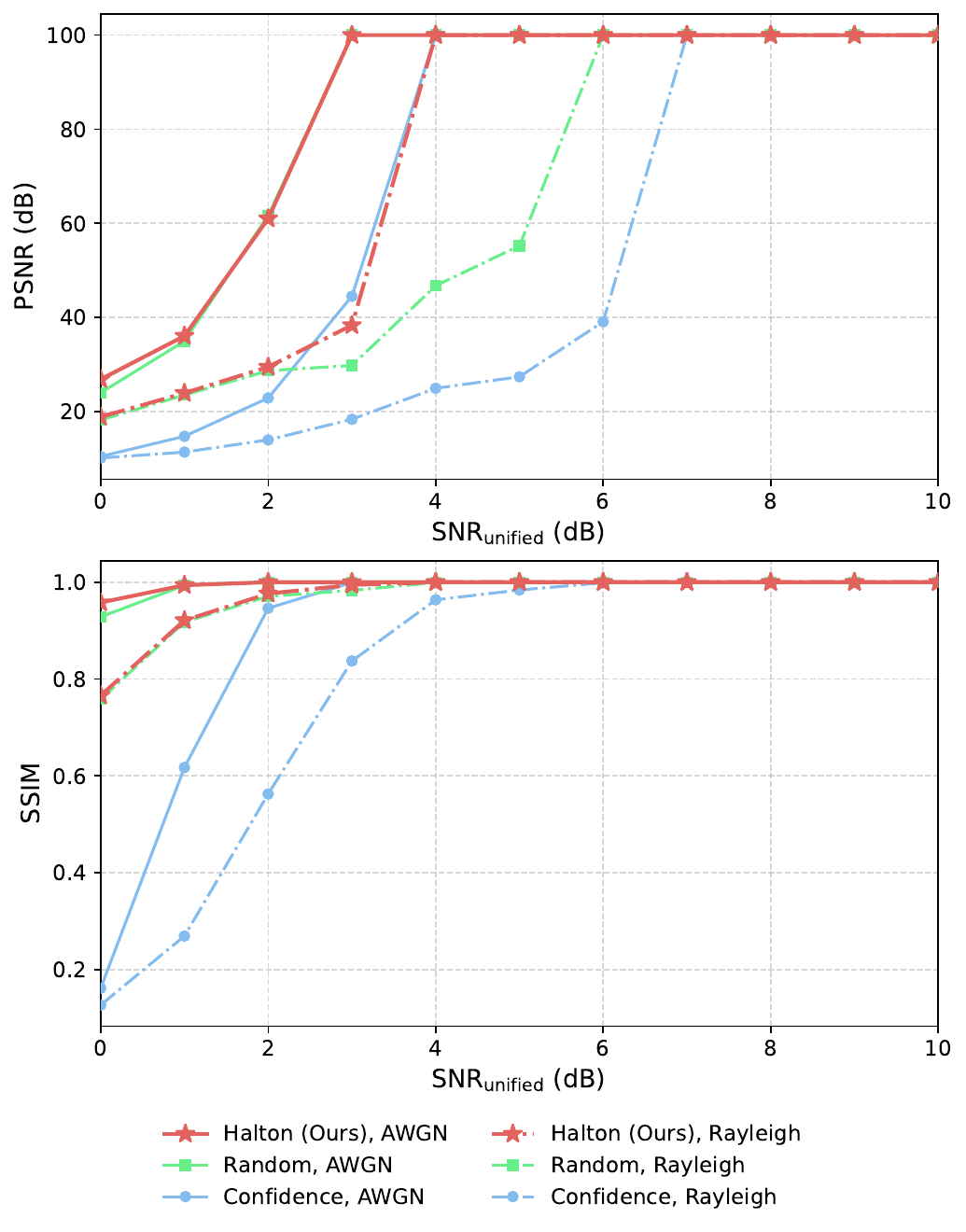}
    \captionsetup{skip=2pt} 
    \caption{Ablation study of the denoising-position selection rule on CIFAR10.}
    \label{fig:ablation_halton}
\end{figure}

\begin{figure}[!tb]
    \centering
    \includegraphics[width=.875\columnwidth]{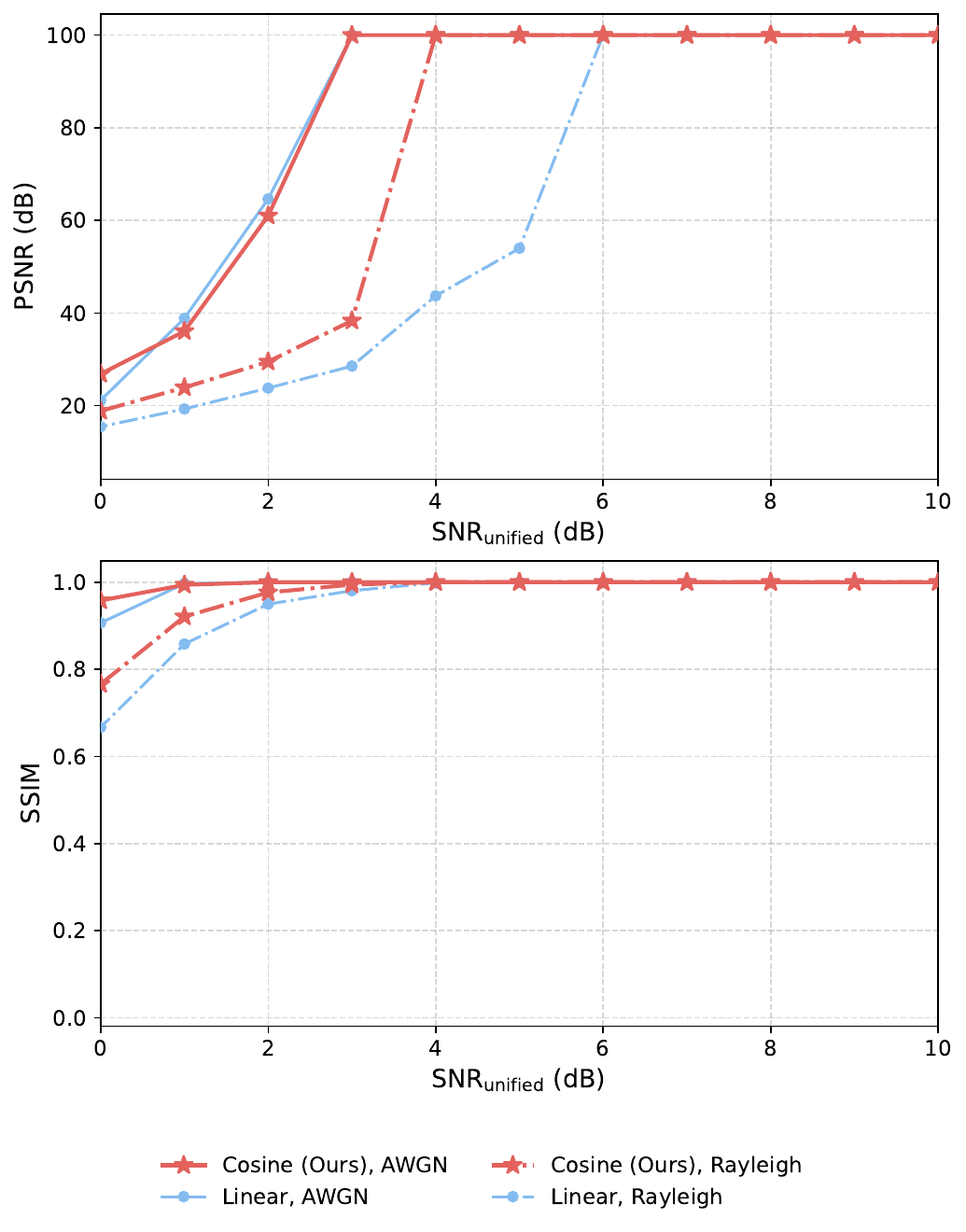}
    \captionsetup{skip=2pt} 
    \caption{Ablation study of the reverse denoising schedule on CIFAR10.}
    \label{fig:ablation_cosine}
\end{figure}
The qualitative results in Figs.~\ref{fig:visual_div2k} and~\ref{fig:visual_Kodak} are consistent with the quantitative curves. 
At $\text{SNR}_{\text{unified}}=0$ dB, DDM-SSCC is already close to perfect reconstruction: the image is visually almost lossless. 
By contrast, DLPR+ECCT reconstructs the early part of the image well, but once bit errors appear, its lossy-plus-residual decoding mechanism causes the subsequent patches to fail, which leads to the severe corruption visible on the right side of the reconstructed image. 
SparseSBC and DeepJSCC still provide visually acceptable reconstructions, with SparseSBC being clearly better; however, both remain far from lossless recovery in fine textures and local details.

\subsubsection{Impact of Denoising Steps} Fig.~\ref{fig:diffu_step_comparison} shows the performance--complexity tradeoff controlled by tuning the number of diffusion steps. 
More denoising steps ($T$) generally improve source modeling and move the exact-recovery threshold to lower $\text{SNR}_{\text{unified}}$ values under both AWGN and Rayleigh fading channels. 
The $T=50$ setting performs best, while the aggressive $T=5$ setting still maintains competitive quality with much lower decoding complexity.

\subsubsection{Contribution of Halton-Guided Denoising} 
Fig.~\ref{fig:ablation_halton} compares Halton-guided denoising with random and confidence-driven position selection. 
Halton-guided denoising reaches the exact-recovery region earlier and provides better pre-threshold PSNR/SSIM, especially under Rayleigh fading. 
This verifies that the denoising order affects arithmetic-coding probability quality, and that the deterministic low-discrepancy order can provide more spatially dispersed context anchors.

\subsubsection{Effectiveness of Cosine denoising schedule} 
Fig.~\ref{fig:ablation_cosine} validates the cosine denoising schedule. 
Compared with the linear schedule, it achieves a lower exact-recovery threshold under both channel models, with a clearer advantage under Rayleigh fading. 
This confirms that conservative early updates and more aggressive late updates improve probability reliability while preserving jump-step efficiency.

\subsubsection{Ablation Studies}

Table~\ref{tab:ablation} further validates the component-wise effectiveness of Halton-guided denoising, cosine scheduling, and mask-ratio-aware calibration. 
The three modules respectively improve spatial coverage, adapt the denoising pace to context reliability, and mitigate stage-dependent probability mismatch. 
All variants share the same high-SNR lossless ceiling, while their pre-threshold differences reflect the reliability of source-bitstream delivery under the fixed channel-protection backend.

\begin{table}[!t]
\centering
\caption{Ablation study of Halton sampling, cosine scheduling, and calibration.}
\label{tab:ablation}
\renewcommand{\arraystretch}{1.25}
\setlength{\tabcolsep}{5pt}
\begin{tabular}{ccc cc}
\toprule
\multicolumn{3}{c}{Components} 
& \multicolumn{2}{c}{AWGN, $\text{SNR}_{\text{unified}}=0$ dB} \\
\cmidrule(lr){1-3}
\cmidrule(lr){4-5}
Halton & Cosine & Calib.
& PSNR (dB) $\uparrow$ & SSIM $\uparrow$ \\
\midrule
$\times$ & $\times$ & $\times$ & 19.35 & 0.8739 \\
$\times$ & $\times$ & $\checkmark$ & 22.33 & 0.9104 \\
$\times$ & $\checkmark$ & $\times$ & 24.42 & 0.9359 \\
$\times$ & $\checkmark$ & $\checkmark$ & 24.00 & 0.9283 \\
$\checkmark$ & $\times$ & $\times$ & 22.28 & 0.9202 \\
$\checkmark$ & $\times$ & $\checkmark$ & 21.15 & 0.9067 \\
$\checkmark$ & $\checkmark$ & $\times$ & 25.30 & 0.9462 \\
$\checkmark$ & $\checkmark$ & $\checkmark$ & \textbf{26.85} & \textbf{0.9581} \\
\bottomrule
\end{tabular}
\end{table}

\section{Conclusion}
\label{sec:conclusion}

This paper has studied lossless pixel-level image transmission from a beyond-semantics perspective.
We have proposed DDM-SSCC, a discrete-diffusion-model-based SSCC framework that turns bidirectional masked restoration into an arithmetic-coding-compatible source codec.
Beyond the basic synchronized diffusion codec, we further clarify the codelength and mutual-information perspective of jump-step denoising, and develop Halton-guided position selection, a mask-ratio-aware cosine schedule, and temperature calibration for diffusion source coding.
These modules respectively reduce confidence-induced spatial clustering, match the denoising pace to context reliability, and alleviate stage-dependent probability miscalibration in the reverse process.
Experiments on CIFAR10, DIV2K-LR-X4, and Kodak over AWGN and Rayleigh fading channels show that DDM-SSCC consistently provides a more favorable source-channel operating point than representative lossless, autoregressive, and semantic baselines, with earlier entry into the exact-recovery regime and stronger pre-threshold fidelity.
Ablation studies further confirm the component-wise benefits of the proposed denoising order, schedule, and calibration strategy.
These results suggest that discrete diffusion is a promising direction for future pixel-level communication systems and can serve as a stronger foundation for subsequent patch-level or higher-level extensions.

\bibliographystyle{IEEEtran}
\bibliography{reference}

\vfill

\end{document}